\documentstyle[sprocl,epsf]{article}

 \catcode`@=11
\def\@cite#1#2{\unskip\nobreak\relax
    \def\@tempa{$\m@th^{\hbox{\the\scriptfont0 #1}}$}%
    \futurelet\@tempc\@citexx}
\def\@citexx{\ifx.\@tempc\let\@tempd=\@citepunct\else
    \ifx,\@tempc\let\@tempd=\@citepunct\else
    \let\@tempd=\@tempa\fi\fi\@tempd}
\def\@citepunct{\@tempc\edef\@sf{\spacefactor=\the\spacefactor\relax}\@tempa
    \@sf\@gobble}
\catcode`@=12

\def\lsim{\mathrel{\raise.2ex\hbox{$<$}\hskip-.8em\lower.9ex\hbox{$\sim$}}}
\def\gsim{\mathrel{\raise.2ex\hbox{$>$}\hskip-.8em\lower.9ex\hbox{$\sim$}}}

\def\abstracts#1{
\begin{center}
{\begin{minipage}{4.7truein}    
                 \footnotesize
                 \parindent=0pt #1\par
                 \end{minipage}}\end{center}
                 \vskip 2em \par}

\begin{document}
\thispagestyle{empty}

\font\fortssbx=cmssbx10 scaled \magstephalf
\hbox to \hsize{
\hskip.1in \raise.02in\hbox{\fortssbx University of Wisconsin - Madison}
\hfill$\vcenter{\small\hbox{\bf MADPH-97-990}
                \hbox{March 1997}}$ }

\vskip3mm

\title{\uppercase{The Search for the Source of the\\
 Highest Energy Cosmic Rays}~\footnote{Talk presented at the International Workshop ``New Worlds in Astroparticle Physics\rlap," Faro, Portugal, Sept.~8--10, 1996.}}
\author{\vskip-1.8em\uppercase{F. Halzen}}
\address{Department of Physics, University of Wisconsin, Madison, WI 53706, USA}
\maketitle

\abstracts{
Active galaxies and gamma ray bursts are the sources of the highest energy photons detected by astronomical telescopes. We speculate that they may be the sources of the highest energy cosmic rays. This makes them true proton accelerators, where the highest energy photons are the decay products of neutral pions photoproduced when the proton beams interacts with ambient radiation. Neutrinos from the decay of charged pions represent an incontrovertible signature for proton acceleration. A main theme of this talk is that their fluxes can be estimated from the measured gamma ray luminosity by model-independent methods, based on dimensional analysis and textbook particle physics.}

\section*{Introduction}
\unskip\smallskip
Cosmic rays form an integral part of our galaxy. Their energy density is qualitatively similar to that of photons, electrons and magnetic fields. It is believed that most were born in supernova blast waves. Their energy spectrum can be understood, up to perhaps 1000\,TeV, in terms of acceleration by supernova shocks exploding into the interstellar medium of our galaxy. Although the slope of the cosmic ray spectrum abruptly increases at this energy, particles with energies in excess of $10^{8}$\,TeV have been observed and, cannot be accounted for by this mechanism. The break in the spectrum, usually referred to as the ``knee\rlap,'' can be best exhibited by plotting the flux multiplied by an energy dependent power $E^{2.75}$; see Fig.~1. 

\begin{figure}[t]
\centering
\hspace{0in}\epsfxsize=3.5in\epsffile{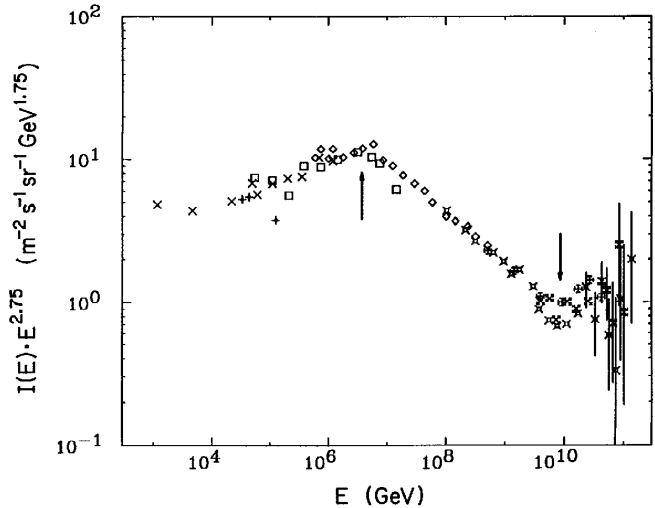}

\caption{Flux of high energy cosmic rays after multiplication by a factor $E^{2.75}$. Arrows point at structure in the spectrum near 1~PeV, the ``knee\rlap," and 10~EeV, the ``ankle\rlap."}
\end{figure}

The failure of supernovae to accelerate cosmic rays above 1000\,TeV energy can be essentially understood on the basis of dimensional analysis. It is sensible to assume that, in order to accelerate a proton to energy $E$, the size $R$ of the accelerator must be larger than the gyroradius of the particle in the accelerating field $B$:
\begin{equation}
R > R_{\rm gyro} = {E\over B} \,.
\end{equation}
This yields a maximum energy
\begin{equation}
E < BR, 
\end{equation}
or,
\begin{equation}
\left[ E_{\rm max}\over 10^5\,\rm TeV \right] = 
\left[ B \over 3\times 10^{-6}\rm G \right] \left[ R\over 50\rm\,pc \right]c\;.
\end{equation}
Therefore particles moving at the speed of light $c$ reach energies up to a maximum value $E_{\rm max}$ which must be less than $10^5$\,TeV for the typical values of $B$ and $R$ for a supernova shock. Realistic modelling introduces inefficiencies in the acceleration process and, yields a maximum energy which is typically two orders of magnitude smaller than the value obtained by dimensional analysis. One therefore identifies the ``knee" with the sharp cutoff associated with particles accelerated by supernovae.

Cosmic rays with energy in excess of $10^{20}$~eV have been observed, some five orders of magnitude in energy above the supernova cutoff. Where cosmic rays with energy in excess of 1000\,TeV are accelerated undoubtedly represents the most urgent problem in cosmic ray astrophysics and, one of the oldest unresolved puzzles in astronomy. In order to beat the dimensional argument just presented, one has to accelerate particles over larger distances $R$, or identify higher magnetic fields $B$. This leads to a variety of speculations\cite{watson}. The large $B$-fields in pulsars, near black holes and in the cores of active galactic nuclei, where they extend over large distances, have been exploited in models where the highest energy cosmic rays originate in point sources. 

Although imaginative arguments actually do exist to avoid this conclusion, it is generally believed that our galaxy is too small and its magnetic field too weak to accelerate the highest energy cosmic rays. Those with energy in excess of $10^{19}$~eV have gyroradii larger than our galaxy and should point back at their sources. Their arrival directions fail to show any correlation to the galactic plane, suggesting extra-galactic origin. As far as extra-galactic sources is concerned, nearby active galactic nuclei distant by $\sim100$~Mpc are obvious candidates. The jets of blazars support magnetic fields of 10~G over distances of $10^{-2}$~parsecs or more. Using Eq.~(3) we reach energies of $10^{20}$~eV, possibly higher because of beaming. Such speculations are reinforced by the fact that a cursory glance at the EGRET and Whipple results is sufficient to convince oneself that blazars are also the dominant (exclusive?) sources of the highest energy gamma rays\cite{whipple}.

More ``mundane" objects such as neutron stars and solar mass black holes may accelerate particle to $10^{20}$~eV in the most extreme circumstances: when a pair of them merges to produce a gamma ray burst. We return to this possibility at the end of this lecture.

\section{Active Galaxies as Proton Accelerators}

AGN are the brightest sources in the Universe. Their engines must not only be
powerful, but extremely compact because their high energy luminosities are
observed to flare by over an order of magnitude over time periods as short as a
day\cite{variability}. Only sites in the vicinity of black holes, a billion
times more massive than our sun, can possibly satisfy the constraints of the
problem. Highly relativistic and confined jets of particles are a common
feature of these objects. It is anticipated that beams, accelerated near the
black hole, are dumped on the radiation in the galaxy which consists of mostly
thermal photons with densities of order 10$^{14}$/cm$^3$; see Fig.~2. The multi-wavelength
spectrum, from radio waves to TeV gamma rays, is produced in the interactions
of the accelerated particles with the magnetic fields and ambient light in the
galaxy. In the more conventional electron models, the highest energy photons
are produced by Compton scattering of accelerated electrons on thermal UV
photons which are scattered from 10~eV up to TeV energy\cite{dermer}. The
energetic gamma rays will subsequently lose energy by electron pair production
in photon-photon interactions with the radiation field of the jet or the
galactic disk. An electromagnetic cascade is thus initiated which, via pair
production on the magnetic field and photon-photon interactions, determines the
emerging gamma-ray spectrum at lower energies. The lower energy photons,
observed by conventional astronomical techniques, are, as a result of the
cascade process, several generations removed from the primary high energy
beams.

\begin{figure}[t]
\centering
\epsfxsize=2.65in\hspace{0in}\epsffile{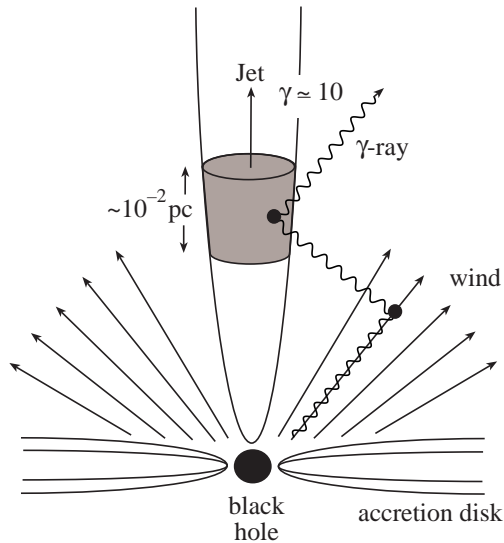}

\caption{Possible blueprint for the production of high energy photons and
neutrinos near the super-massive black hole powering an AGN. Particles,
accelerated in sheet like bunches moving along the jet, interact with photons
radiated by the accretion disk or produced by the interaction of the
accelerated particles with the magnetic field of the jet.}
\end{figure}

\break
The EGRET instrument on the Compton Gamma Ray Observatory has detected high
energy gamma-ray emission, in the range 20 MeV--30 GeV, from over 100
sources\cite{EGRET}. Of these sources 16 have been tentatively, and 42 solidly
identified with radio counterparts. All belong to the ``blazar" subclass,
mostly Flat Spectrum Radio Quasars, while the rest are BL-Lac
objects\cite{Mattox}. In a unified scheme of AGN, they correspond to Radio Loud
AGN viewed from a position illuminated by the cone of a relativistic
jet\cite{padovani}. Moreover of the five TeV gamma-ray emitters identified by
the air Cherenkov technique, three are extra-galactic and are also nearby
BL-Lac objects\cite{whipple}. The data therefore strongly suggests that the
highest energy photons originate in jets beamed to the observer. Several of the
sources observed by EGRET have shown strong variability, by a factor of 2 or so
over a time scale of several days\cite{variability}. Time variability is more
spectacular at higher energies. On May 7, 1996 the Whipple telescope observed
an increase of the TeV-emission from the blazar Markarian 421 by a factor 2 in
1 hour reaching, eventually, a value 50 times larger than the steady flux. At
this point the telescope registered 6 times more photons from the Markarian
blazar, more distant by a factor $10^5$, than from the Crab supernova
remnant\cite{421flare}.

Does pion photoproduction by accelerated protons play a central role in blazar
jets? This question has been extensively debated in recent years\cite{pic}. If
protons are accelerated along with electrons, they will acquire higher
energies, reaching PeV--EeV energy because of reduced energy losses. High
energy photons result from proton-induced photoproduction of neutral pions on
the ubiquitous UV thermal background. Accelerated protons thus initiate a
cascade which dictates the features of the spectrum at lower
energy\cite{biermann}. From a theorist's point of view the proton blazar has
attractive features. Protons, unlike electrons, efficiently transfer energy
from the black hole in the presence of the high magnetic fields required to
explain the confinement of the jets\cite{mannheimB}. Protons provide a
``natural'' mechanism for energy transfer from the central engine over
distances as large as 1~parsec, as well as for the observed heating of the
dusty disk over distances of several hundred parsecs\cite{biermann}. More to
the point, the issue of proton acceleration can be settled experimentally
because the proton blazar is a source of high energy protons and neutrinos, not
just gamma rays\cite{PR}.

Weakly interacting neutrinos can, unlike high energy gamma-rays and high energy
cosmic rays, reach us from more distant and much more powerful AGN. It is
likely that absorption effects explain why Markarian 421, the closest blazar on
the EGRET list at a distance of $\sim$~150~Mpc , produces the most prominent
TeV signal. Although the closest, it is one of the weakest; the reason that it
is detected whereas other, more distant, but more powerful, AGN are not, must
be that the TeV gamma rays suffer absorption in intergalactic space by
interaction with background infra-red light\cite{salomon}. This most likely
provides the explanation why much more powerful quasars with significant high
energy components such as 3C279 at a redshift of 0.54 have not been identified
as TeV sources.
Undoubtedly, part of the TeV flux is also absorbed on the infrared light in the
source; we will return to this further on.

First order Fermi acceleration offers a very attractive model for acceleration
in jets, providing, on average, the right power and spectral shape. A cosmic
accelerator in which the dominant mechanism is first order diffusive shock
acceleration, will indeed produce a spectrum
\[
dN/dE\propto E^{-\gamma} \,, \label{eq:dN/dE}
\]
with $\gamma \sim 2 +\epsilon$, where $\epsilon$ is a small number. For strong
ultra-relativistic shocks it can be negative ($\sim -0.3$ ). Confronted with
the challenge of explaining a relatively flat multi-wavelength photon emission
spectrum which extends to TeV energy, models have converged on the blazar
blueprint shown in Fig.~2. Particles are accelerated by Fermi shocks in bunches
of matter travelling along the jet with a bulk Lorentz factor of order $\gamma
\sim 10$. Ultra-relativistic beaming with this Lorentz factor provides the
natural interpretation of the observed superluminal speeds of radio structures
in the jet\cite{rees}. In order to accommodate bursts lasting a day in the
observer's frame, the bunch size must be of order $\Gamma c \Delta t \sim
10^{-2}$~parsecs. Here $\Gamma$ is the Doppler factor, which for observation
angles close to the jet direction is of the same order as the Lorentz
factor~\cite{padovani}. These bunches are, in fact, more like sheets, thinner
than the jet's width of roughly 1~parsec. The observed radiation at all
wavelengths is produced by the interaction of the accelerated particles in the
sheets with the ambient radiation in the AGN, which has a significant component
concentrated in the so-called ``UV-bump\rlap".

In electron models the multi-wavelength spectrum consists of  three components:
synchrotron radiation produced by the electron beam on the $B$-field in the
jet, synchrotron photons Compton scattered to high energy by the electron beam
and, finally, UV photons Compton scattered by the electron beam to produce the
highest energy photons in the spectrum\cite{dermer}. The seed photon field can
be either external, e.g.\ radiated off the accretion disk, or result from the
synchrotron radiation of the electrons in the jet, so-called
synchrotron-self-Compton models. The picture has a variety of problems. In
order to reproduce the observed high energy luminosity, the accelerating
bunches have to be positioned very close to the black hole. The photon target
density is otherwise insufficient for inverse Compton scattering to produce the
observed flux. This is a balancing act, because the same dense target will
efficiently absorb the high energy photons by $\gamma\gamma$ collisions. The
balance is difficult to arrange, especially in light of observations showing
that the high energy photon flux extends beyond TeV energy\cite{whipple}. The
natural cutoff occurs in the 10--100~GeV region\cite{dermer}. Finally, in order
to prevent the electrons from losing too much energy before producing the high
energy photons, the magnetic field in the jet has to be artificially adjusted
to less than 10\% of what is expected from equipartition with the radiation
density.

For these, and the more general reasons already mentioned in the introduction,
the proton blazar has been developed. In this model protons as well as
electrons are accelerated. Because of reduced energy loss, protons can produce
the high energy radiation further from the black hole. The more favorable
production-absorption balance far from the black hole makes it relatively easy
to extend the high energy photon spectrum above 10 TeV energy, even with bulk
Lorentz factors that are significantly smaller than in the inverse Compton
models. Two recent incarnations of the proton blazar illustrate that these
models can also describe the multi-wavelength spectrum of the
AGN\cite{mannheim,protheroe}. Because the seed density of photons is still much
higher than that of target protons, the high energy cascade is initiated by the
photoproduction of neutral pions by accelerated protons on ambient light via
the $\Delta$ resonance. The protons collide either with synchrotron photons
produced by electrons\cite{mannheim}, or with the photons radiated off the
accretion disk\cite{protheroe}, as shown in Fig.~2.

\section{The Neutrino Flux from Blazar Jets}

Model-independent evidence that AGN are indeed cosmic proton accelerators can
be obtained by observing high energy neutrinos from the decay of charged pions,
photoproduced on the $\Delta$ resonance along with the neutral ones. The
expected neutrino flux can be estimated in six easy steps.

\begin{enumerate}

\item

The size of the accelerator $R$ is determined by the duration, of order 1 day,
over which the high energy radiation is emitted:
\begin{equation}
R=\Gamma t c = 10^{-2}\mbox{ parsecs for }t = 1\rm\ day.
\end{equation}

\item

The magnitude of the $B$-field can be calculated from equipartition with the
electrons, whose energy density is measured experimentally:
\begin{equation}
{B^2 \over 2 \mu_0}  = \rho\rm (electrons) \,{\sim} \, 1\ erg/cm^3.
\end{equation}
This yields a value for the magnetic field of 5~Gauss. A similar value is
obtained by scaling $B$-fields in the jets of Fannaroff-Riley type II galaxies
at kiloparsec distances, to the Markarian 421 luminosity, and to transverse
distances in the milliparsec range\cite{mannheimB}.

\item

In shock acceleration the gain in energy occurs gradually as a particle near
the shock scatters back and forth across the front gaining energy with each
transit. The proton energy is limited by the lifetime of the accelerator and
the maximum size of the emitting region\cite{PR} $R$,
\begin{equation}
E  < K Z e B R c\,.
\label{eq:Emax}
\end{equation}
Here $Ze$ is the charge of the particle being
accelerated and $B$
the ambient magnetic field. The upper limit basically follows from dimensional
analysis. It can also be derived from the simple requirement that the
gyroradius of the accelerated particles must be contained within the
accelerating region $R$. The numerical constant $K\sim 0.1$ depends on the
details of diffusion in the vicinity of the shock, which determine the
efficiency by which power in the shock is converted into acceleration of
particles. In some cases it can reach values close to 1. The maximum energy
reached is
\[
E_{\rm max} = e B R c = 5 \times 10^{19} \rm\ eV
\]
for $B = 5$~Gauss and  $R = 0.02$~parsecs. We here assumed that the boost of
the energy in the observer's frame approximately compensates for the efficiency
factor, i.e.\ $K \,\Gamma \sim 1$.

The neutrino energy is lower by two factors which take into account i) the
average momentum carried by the secondary pions relative to the parent proton
($\left< x_F\right> \simeq 0.2$) and ii) the average energy carried by the
neutrino in the decay chain $\pi^+ \rightarrow \nu_\mu \mu^+ \rightarrow  e^+
\nu_e \bar{\nu}_\mu$, which is roughly 1/4 of the pion energy because equal
amounts of energy are carried by the four leptons. The maximum neutrino energy
is
\begin{equation}
E_{\nu\,\rm max} = E_{\rm max} \left< x_F\right> {1\over4} \simeq 10^{18}\rm\,
eV\,,
\end{equation}
i.e.\ neutrinos reach energies of $10^3$~PeV.

\item

The neutrino spectrum can now be calculated from the observed gamma ray
luminosity. We recall that approximately equal amounts of energy are carried by
the four leptons that result from the  decay chain $\pi^+ \rightarrow \nu_\mu
\mu^+ \rightarrow e^+ \nu_e \bar{\nu}_\mu$. In addition the cross sections for
the processes $p\gamma \rightarrow p \pi^0$ and  $p\gamma \rightarrow n \pi^+ $
at the $\Delta$ resonance are in the approximate ratio of $2:1$. Thus 3/4 of
the energy lost to photoproduction ends up in the electromagnetic cascade and
1/4 goes to neutrinos, which corresponds to a ratio of neutrino to gamma
luminosities ($L_\nu : L_\gamma$) of $1:3$. This ratio is somewhat reduced when
taking into account that some of the energy of the accelerated protons is lost
to direct pair production ($p +\gamma\rightarrow e^+ e^- p$):
\begin{equation}
L_\nu\, =\,\frac{3}{13}L_\gamma\,.
\end{equation}
In order to convert above relation into a neutrino spectrum we have to fix the
spectral index. We will assume that the target photon density spectrum is
described by a $E^{-(1+\alpha)}$ power law, where $\alpha$ is small for AGN
with flat spectra. The number of target photons above photoproduction threshold
grows when the proton energy $E_p$ is increased. If the protons are accelerated
to a  power law spectrum with spectral index $\gamma\,(=2+\epsilon)$, the
threshold effect implies that the spectral index of the secondary neutrino flux
is also a power law, but with an index flattened by $(1+\alpha)$ as a result of
the increase in target photons at resonance when the proton energy is
increased:
\begin{equation}
{dN_\nu\over dE_{\nu}} = {\cal N}
\left[
{ E_{\nu} \over E_{\nu\rm\,max} }
\right] ^{-(1+\epsilon-\alpha)}.
\end{equation}
For a standard non relativistic shock with $\epsilon = 0$ and a flat photon
target with $\alpha=0$, the neutrino spectrum will flatten by just one unit
giving $E{dN_\nu\over dE} \sim \rm constant$. From Eqs.~(5) and (6)
\begin{equation}
\int^{E_{\nu\rm\,max}} E{dN_\nu\over dE_{\nu}} dE_{\nu} \simeq
{\cal N} {E_{\nu\rm\,max}^2 \over 1-\epsilon+\alpha} \simeq {3\over13}L_\gamma\,.
\end{equation}
The calculation is stable as long as $\epsilon - \alpha$ is smaller than 1
because the luminosity integral is not sensitive to the lower limit of the
integration.

\item

Assuming that the high energy $\gamma$ ray flux from Markarian 421 results from
cascading of the gamma ray luminosity produced by Fermi accelerated protons, we
obtain the neutrino flux from the measured value\cite{whipple} of $L_{\gamma}$
of $2 \times 10^{-10}$~TeV~cm$^{-2}$~s$^{-1}$:
\begin{equation}
{dN_\nu\over dE_{\nu}} = {3\over13}{L_\gamma\over E_{\nu\rm\,max}} \;{1- \epsilon +
\alpha \over E_{\nu}} \left[ { E_{\nu} \over E_{\nu\rm\,max} }
\right]^{-(\epsilon+\alpha)} \sim
{5 \times 10^{-17}\,{\rm cm^{-2}\,s^{-1}}\over E_{\nu}} \,,
\end{equation}
where the numerical estimate corresponds to $\alpha=\epsilon=0$ and the value
of $E_{\nu\rm\,\max}$ of Eq.~(4). This calculation reveals that for the small
values of $\epsilon$ and $\alpha$ anticipated, the neutrino flux is essentially
determined by the value for $E_{\nu\rm\,\max}$.

\item

In order to calculate the diffuse flux from the observed blazar distribution,
we note that the EGRET collaboration has constructed a luminosity function
covering the observation of the $\sim$20 most energetic blazars and estimated
the diffuse gamma ray luminosity\cite{chiang}. From the ratio of the diffuse
gamma ray flux and the flux of Markarian 421, we obtain that the effective
number of blazars with Markarian 421 flux is $\sim 130\rm~sr^{-1}$. The
diffuse neutrino flux is now simply estimated by multiplying the calculated
flux for Markarian 421 by this factor. A correction for the difference in
spectral indices of gamma ray and neutrino fluxes enhances the neutrino flux by
a factor of three. The flux corresponds to an energy regime well below the high
energy cut-off. The transition to the cutoff should be smooth because of the
superposition of the different redshifts and cut-off energies of the individual
blazars.
\end{enumerate}

This concludes our calculation. It illustrates how the proton blazar, unlike
the electron blazar, requires no large Doppler factors and no fine-tuning of
parameters. For the proton blazar, radiation and magnetic fields are in
equipartition, the maximum energy matches the $BR$ value expected from
dimensional analysis and, finally, the size of the bunches is similar to the
gyroradius of the highest energy protons. It is not a  challenge to increase
gamma ray energies well beyond the TeV energy range. Reasonable variations of
the values of magnetic field strength $B$, the efficiency parameter $K$ and the
Doppler boost factor $\Gamma$ may allow us to account for the highest energy
cosmic rays with $E\sim 3 \times10^{20}~$eV.

Also, our calculation demonstrates why models\cite{mannheim,protheroe} which
differ in many aspects, yield very similar predictions for the neutrino flux,
consistent with the ones obtained here\cite{hill}; this is illustrated in
Fig.~3.

\begin{figure}[h]
\centering
\epsfxsize=3.75in\hspace{0in}\epsffile{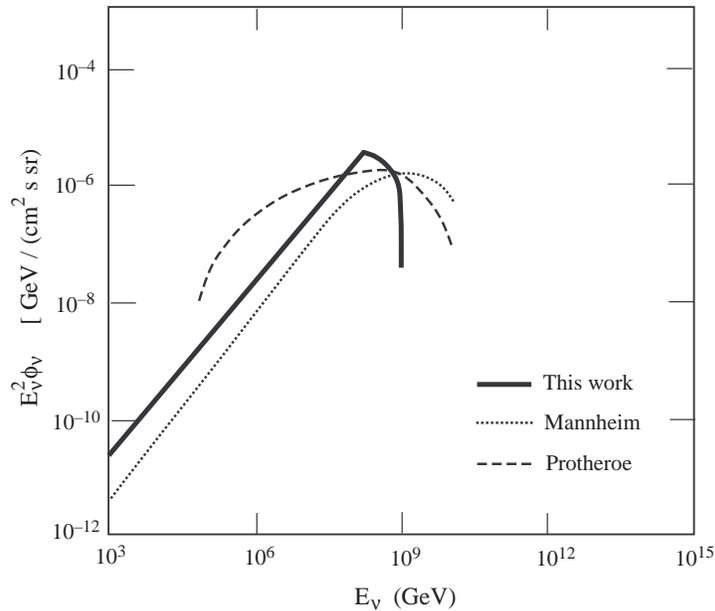}

\caption{Diffuse neutrino flux from blazars. The numerical result of Eq.~(11) multiplied by $3 \times 130$~sr$^{-1}$ and corrected for redshift in the cutoff is compared to recent calculations \protect\cite{mannheim,protheroe}.}
\end{figure}

\section{Cosmic Rays and High Energy Neutrinos}

Rather than scaling to the TeV gamma ray flux, we can use the cosmic ray flux
at ultra high energies to bracket expectations for the neutrino flux. Models
for proton acceleration in hot spots of Fanaroff-Riley type II galaxies can
explain the observed cosmic ray spectrum above $\sim 10^{18}$~eV\cite{rachen}.
This energy corresponds to the ``ankle" in the spectrum, where the observed
spectral index flattens from 3 to 2.7. The model requires an $E^{-2}$, or
flatter, injection spectrum which steepens above $10^{17}$~eV to the observed
$E^{-2.7}$ spectrum as a result of energy loss in the source, interactions with
the microwave background, and cosmological evolution\cite{rachen}. Because of
the strict limitations on the density of target photons at the acceleration
site, previously discussed, roughly similar neutrino and proton luminosities
are expected\cite{PR}. In order to understand this balance it is important to
realize that in astrophysical beam dumps the accelerator and production target
form a symbiotic system. Although larger target density may produce more
neutrinos, it also decelerates the protons producing them, in a delicate
acceleration-absorption balance. Equal cosmic ray and neutrino luminosity
implies:
\begin{equation}
\int dE_{\nu} (E_{\nu} \, dN_\nu/dE_{\nu}) \sim L_{\rm CR} \sim 10^{-9}\rm\ TeV\
cm^{-2}\, s^{-1} \, sr^{-1} \;.
\end{equation}
The bulk cosmic ray luminosity has been conservatively estimated by assuming
that it is due to an $E^{-2.7}$ spectrum above $\sim 10^{17}~$eV. This spectrum
has been normalized to the observed EeV cosmic rays. It is interesting to note
that this luminosity is a factor 25 below the measured diffuse gamma ray
luminosity from AGN\cite{chiang}. This is, within an order of magnitude, in
agreement with the relation of neutrino and gamma ray luminosities of Eq.~(5),
and, if anything, implies a conservative estimate of the neutrino flux.

Assuming a generic $E^{-2}$ neutrino spectrum, the equality of cosmic-ray and
neutrino luminosities implies:
\begin{equation}
E_{\nu}{dN_\nu\over dE_{\nu}} \sim {10^{-10}\over E_{\nu}\,(\rm TeV)} \, \rm
cm^{-2}\ s^{-1}\ sr^{-1} \;. \label{eq:flux}
\end{equation}
A not too different result is obtained by assuming equal numbers of neutrinos
and protons, rather than equal luminosities. It is clear that our estimate is
conservative because the proton flux reaching Earth has not been corrected for
absorption of protons in ambient matter in the source, or in the interstellar
medium.

\section{Event Rates in Underground Muon Neutrino Telescopes}

The probability to detect a TeV neutrino is roughly $10^{-6}$\cite{PR}. It is
easily computed from the requirement that, in order to be detected, the
neutrino has to interact within a distance of the detector which is shorter
than the range of the muon it has produced. In other words, in order for the
neutrino to be detected, the produced muon has to reach the detector.
Therefore,
\begin{equation}
P_{\nu\to\mu} \simeq {R_\mu\over \lambda_{\rm int}} \simeq A E_{\nu}^n \,,
\end{equation}
where $R_{\mu}$ is the muon range and $\lambda_{\rm int}$ the neutrino
interaction length. For energies below 1~TeV, where both the range and cross
section depend linearly on energy, $n=2$. Between TeV and PeV energies $n=0.8$
and $A=10^{-6}$, with $E$ in TeV units. For EeV energies $n=0.47$, $A =10^{-2}$
with $E$ in EeV.

We are now ready to compute the diffuse neutrino event rate by folding the
neutrino spectrum of Eq.~(8) with the detection probability of Eq.~(11). We
also multiply by 130~sr$^{-1}$ for the effective number of sources:
\begin{equation}
\phi^\nu = \int^{E_{\nu\,\rm max}} {dN_\nu\over dE_{\nu}} P_{\nu\to\mu}(E_{\nu}) dE_{\nu} \simeq
40\rm\ km^{-2}\,year^{-1}\,sr^{-1} \,.
\end{equation}
which implies a yield of two neutrinos every three days in a kilometer-scale detector, assuming only $2 \pi$ coverage.

The steeper, but lower luminosity, flux of Eq.~(10) predicts more events when
folded with Eq.~(11), about $150\rm\ km^{-2}\,year^{-1}\,sr^{-1}$ assuming that
the flux extends down to TeV energy. The result does not depend strongly on the
lower limit of the neutrino integral, it only drops by a factor of three if the
neutrino flux flattens below 100~TeV. We again conclude that a kilometer-scale
neutrino detector may be required\cite{halzen}. It is however important to
realize that, had we assumed a $E^{-1}$ spectrum, the resulting flux would have
scaled with the ratio of luminosities to about an order of magnitude below
Eq.~(12). The energy dependence of the detection efficiency of  underground
muon neutrino detectors is such that most of the events are detected in the
high (low) energy end for a $E^{-1}$ ($E^{-2}$) spectrum.

\section{Evidence for the Proton Blazar?}

Astronomy with protons becomes possible once their energy has reached a value
where their gyroradius in the microgauss galactic field exceeds the dimensions
of the galaxy. Provided that intergalactic magnetic fields are not too strong,
protons with $10^{20}$~eV energy point at their sources with degree-accuracy.
At this energy their mean-free-path in the cosmic microwave background is
unfortunately reduced to only tens of megaparsecs. A clear window of
opportunity emerges: Are the directions of the cosmic rays with energy in
excess of  ${\sim} 5 \times 10^{19}$~eV correlated to the nearest AGN
(red-shift $z$ less than 0.02), which are known to be clustered in the
so-called ``super-galactic" plane? Although far from conclusive, there is some
evidence that such a correlation may exist\cite{stanev}. Lack of statistics at
the highest energies is a major problem. Future large aperture cosmic ray
detectors such as the new Utah HIRES air fluorescence detector and the Auger
giant air shower array will soon remedy this aspect of the
problem\cite{watson}.

Another problem is that the pointing accuracy is not really understood. It depends on the distance $d$ to the source and the gyroradius in the intergalactic magnetic field:
\begin{equation}
\theta \cong {d\over R_{\rm gyro}} = {dB\over E} \,,
\end{equation}
or scaled to units relevant to the problem
\begin{equation}
{\theta\over0.1^\circ} \cong { \left( d\over 1{\rm\ Mpc} \right) \left( B\over 10^{-9}{\rm\,G} \right) \over \left( E\over 3\times10^{20}\rm\, eV\right) }\,.
\end{equation}
Speculations on the strength of the inter-galactic magnetic field range from $10^{-7}$ to $10^{-12}$~Gauss. For a distance of 100~Mpc, the resolution may therefore be anywhere from sub-degree to nonexistent. It is reasonable to expect that magnetic fields are higher in regions where matter is clustered. Higher values may therefore be appropriate for the local cluster and the super-galactic plane, regions relevant to this problem. Optimistically, one may anticipate that future high statistics experiment such as HIRES in Utah and the Auger giant air shower array will provide indirect information on the magnitude of the magnetic fields between galaxies.

We have already drawn attention to the 10~TeV maximum photon energy as the
demarkation line between the electron and proton blazars. The $\sim10$~GeV
cutoff in the inverse Compton model can be pushed to the TeV range in order to
accommodate the Whipple data on Markarian 421, but not beyond. Bringing the
accelerator closer to the black hole may yield photons in excess of 10~TeV
energy --- they have, however, no chance of escaping without energy loss on the
dense infrared background at the acceleration site. HEGRA has been monitoring
the 10 closest blazars, including Markarian 421, with its dual telescope
systems: the scintillator and the naked photomultiplier detector arrays. The 
announcement\cite{rhode} that their upper limit on the photon flux of 50~TeV
and above for the aggregate emission from the ten nearest blazars, may be a
signal, could provide the first compelling evidence that blazar jets are indeed
proton accelerators.

In summary, there are hints that active galaxies may be true particle
accelerators with proton beams dictating the features of the spectrum. With the
rapidly expanding Baikal and AMANDA detectors producing their first hints of
neutrino candidates\cite{domogatsky,hulth}, observation of neutrinos from AGN
would establish the production of pions and identify the acceleration of
protons as the origin of the highest energy photons. A definite answer may not
be known until these detectors reach kilometer size. Neutrino telescope
builders should take note that, although smaller neutrino fluxes are predicted
than in the generic AGN models of a few years ago\cite{stecker}, they are all
near PeV energy where the detection efficiency is increased and the atmospheric
neutrino background negligible. Because of the beaming of the jets, the
neutrinos have a flatter spectrum peaking near the $10^6$~TeV maximum energy.
The actual event rates are, in the end, not very different.

If confirmed, these models strongly favor the construction of neutrino
telescopes following a distributed architecture, with large spacings of the
optical modules and relatively high threshold\cite{Halzen}. This also opens up
opportunities for alternative techniques such as the radio technique, or the
detection of horizontal air showers with giant air shower arrays\cite{parente}.
Optimists, on the other hand, can find reasons to anticipate the discovery of
AGN neutrinos with much smaller telescopes. With a sufficiently high proton
target density in the acceleration region, much larger fluxes of neutrinos may
be produced in a proton-proton cascade. The predicted fluxes are however
model-dependent\cite{protheroe}. It is also possible, even likely, that
accelerated protons which produce neutrinos do not escape the source, or escape
after significant energy loss. Such absorption effects increase the neutrino
flux relative to the observed high energy cosmic ray flux, also leading to
larger neutrino fluxes.

\section{Comments on the Highest Energy Cosmic Rays: Particles with Macroscopic Energy}

Ever since the pioneering Haverah Park experiment\cite{Watson} revealed that cosmic particles are accelerated up to $10^{20}$~eV energy, the origin of these very highest energy particles has been hotly debated. The recent observation of isolated events with even higher energy is nothing less than paradoxical; they imply aspects of particle physics or astrophysics not revealed in previous experiments.

	In October 1991, the Fly's Eye cosmic ray detector recorded an event of energy $3.0\pm^{0.36}_{0.54}\times 10^{20}$~eV~\cite{flyes}. This event, together with an event recorded by the Yakutsk air shower array in May 1989~\cite{yakutsk}, of estimated energy $\sim2\times 10^{20}$~eV, are the two highest energy cosmic rays ever seen. A recent paper~\cite{akeno} reports that the Akeno Giant Air Shower Array, an instrument of over 100 scintillation detectors spread over a $100$~km$^2$ area in the Kofu vineyards in Japan, also recorded an event of (1.7--$2.6)\times10^{20}$~eV energy. These cosmic rays' energies exceed by a factor of a hundred million those of the highest energy particle beams achieved with man-made accelerators. When colliding with atmospheric nuclei, the center of mass energy is of the order of 700~TeV or $\sim 50$ Joules, almost 50 times that of the future Large Hadron Collider at CERN. 

	Nature accelerates microscopic particles to macroscopic energy by mechanisms that are still a matter of speculation. e already noted that the gyroradius of a $10^{19}$~eV proton in the $3\times10^{-6}$~gauss galactic field is roughly 10,000 light-years, comparable to the size of our galaxy. So, $10^{20}$~eV particles should travel in straight paths from their sources through the galactic and intergalactic magnetic fields. Textbook particle physics is sufficient to derive that $10^{20}$~eV protons travelling through the omnipresent 3K photon background will photoproduce pions, and will thus be demoted in energy over a distance of less than 10~Mpc, i.e.\ much less than the 100~Mpc plus distance from the posited sources. Although one can do astronomy with the highest energy {\em charged\/} particles, their reach into the Universe is very limited by attenuation in the cosmic photon background.

	Here are the facts. The mean free path of a $3\times 10^{20}$~eV proton in the cosmic photon background is only 8.8~Mpc. The probability for a proton to traverse 100~Mpc without an interaction is $1.16\times 10^{-5}$. Alternatively, a cosmic ray proton needs an energy of $3\times 10^{22}$~eV to reach Earth from a 100~Mpc source with the observed energy.  Needless to say, acceleration to $10^{22}$~eV becomes a challenge, even in the context of supermassive black holes~\cite{Sigl}. In the case of the Fly's Eye event no obvious candidate sources exist near or within the error box of the observed arrival direction. (There are some candidate AGN sources for which the distance is, however, unknown). From the previous discussion it is clear that the identification of the highest energy cosmic rays with protons is problematic. The above arguments apply, {\it mutatis mutandis}, to nuclei.

	As previously mentioned, above a threshold of about $5\times10^{19}$~eV, the so-called Greisen-Zatsepin cutoff, cosmic rays lose energy in the universal photon background by photoproducing pions which eventually decay into photons and neutrinos. It has therefore been anticipated that at the highest energies, $\gamma$-rays and neutrinos dominate the spectrum. We already discussed that active galaxies may be powerful neutrino sources. Weakly interacting neutrinos escape acceleration sites without absorption and travel without attenuation through interstellar space. They may form the dominant high energy component of cosmic ray accelerators if the highest energy protons (and photons) are absorbed in the ambient matter in the source or in interstellar space. The proton origin of the high energy events has therefore been questioned~\cite{Sommers,Halzen}. 

	Photons of $10^{20}$~eV energy interact with the Earth's magnetic field by pair-producing electrons and start a cascade well before entering the atmosphere. The highest energy cosmic rays are not photons\cite{Halzen}. Neutrino origin is, on the contrary, consistent with the observed air shower profile. A very high energy muon-neutrino interacting high in the atmosphere transfers roughly one fifth of its energy to an air nucleus which subsequently produces a hadron-like cascade. Such an event would, of course, be atypical since the atmosphere is relatively transparent to the weakly interacting neutrinos. It would be the tip of an iceberg that reaches its maximum inside the Earth, waiting to be discovered by the high energy neutrino telescopes now under construction.

	One may argue that proton and neutrino assignments are equally problematic, by a factor $10^{-5}$. The factor represents the probability of reaching Earth through the microwave background in the case of a proton and the ratio of the neutrino to proton interaction cross section with the atmosphere in the case of neutrino origin. So, the skeptic will conclude that these particles are neither protons, nor photons nor neutrinos, and that nothing accelerates particles to $10^{20}$~eV anyway. This is the paradox. The optimistic theorist may argue that we are observing a new particle, created in a topological defect, efficiently {\em de}celerated from Planck energies. The experimentalist will argue that we need more data. This is quite a challenge given a flux of order $10^{-3}/$(km$^2$ sr year), which points at the need for a detector with an effective area of order 10000~km$^2$ which will be hopefully commissioned as the ``Auger Giant Air Shower Detector\rlap". In the meantime Utah's HIRES has a shot at the same problem.

\section{Are GRBs the Sources of the Highest Energy Cosmic Rays?}

Our speculation that AGN jets are the sources of the highest energy cosmic rays has been very much inspired by the fact that they are also the sources of the highest energy photons. Photons of GeV energy are also produced in gamma ray bursts (GRBs)\cite{meegan} and this raises the possibility that two of the outstanding enigmas in astronomy, high energy cosmic rays and GRBs, may actually represent a single puzzle.

GRBs, typically, release a fraction of a solar mass of energy ($\sim10^{53}$~ergs) over millisecond time scales in photons with a very hard spectrum. Approaching supernova-scale energy release over such short time scales, GRBs are clearly not in the realm of the stars and plasmas of everyday astronomy. They must represent true high energy cosmic accelerators, possibly merging neutron stars or black holes.

As we have seen, similar speculations emerge when faced with the formidable challenge of  identifying the origin of the highest energy cosmic rays. It has been suggested\cite{waxman} that, though unknown, the same cataclysmic mechanism that produces GRBs also produces the highest energy cosmic rays. This association is reinforced by other features of the data besides the phenomenal energy and luminosity of their respective sources:

\begin{itemize}

\item

with improved statistics it has become increasingly suggestive that both GRBs and the highest energy cosmic rays are produced in cosmological sources, i.e.\ distributed throughout the Universe, and

\item

the average rate (over volume and time) at which energy is injected into the Universe as gamma rays from GRBs is similar to the rate at which energy is injected in ${>}10^{20}$~eV cosmic rays in order to produce the observed cosmic ray flux.

\end{itemize}

The obvious question is whether GRBs and cosmic rays are actually observed in coincidence? As noted above, the high energy protons lose energy on the universal background of cosmic photons, the highest energy cosmic rays observed at Earth reach us from no further than a few tens of megaparsecs. It is an observational fact that the rate of GRBs within this limited cosmological volume is only one every few hundred years, while $10^{20}$~eV cosmic rays are detected at a rate of one every few years. There is no contradiction here! As a result of deflection and energy dispersion of the cosmic rays by random magnetic field the source pulse is time broadened by non-straight, lengthened paths. Intergalactic magnetic fields of order $10^{-12}$ Gauss are sufficient to spread the cosmic ray signal over 100 years. This, unfortunately, implies that one does not expect a sharp correlation between the observed directions and arrival times unless GRB are not truly cosmological and their concentration is somehow ``locally" enhanced within a radius of tens of megaparsecs.

Independently of the magnitude of the magnetic fields in the source and in the interstellar medium, sharp coincidences between photons and charged cosmic rays are excluded because of propagation effects inside our own galaxy. The deflection angle of a $10^{20}$~eV cosmic ray is expected to be of order 1 degree as a result of propagation over kiloparsec distances in the microgauss magnetic field of our galaxy. The angle is inversely proportional to the energy of the particle. The wriggling of the trajectory of the cosmic ray by the magnetic field produces delay in its arrival of order several months, scaling inversely with the square of the energy. As previously discussed, cosmic ray times are expected to have been spread by hundreds of years rather than months. This has not discouraged the search for possible coincidences. The Utah Fly's Eye detector observed the highest energy cosmic ray ever on October 15, 1991. Within the error box of its direction the BATSE detector on the Compton Gamma Ray telescope observed GRB (910503) 5.5 months earlier. This event happened to be the highest fluency event in the BATSE catalogue. This burst also had photons with energy as high as 10 GeV. With the usual pitfalls of post-factum statistics it is difficult to judge the weight of evidence of such association. Is this an atypical nearby event? Are the GRBs not truly cosmological? Is this association fortuitous? The future will tell. Other, less dramatic associations of GRBs and high energy cosmic rays have been suggested. Independently of the polemic, cosmic rayers will undoubtedly look out for future events in the direction of the February 17, 1994 ``Superbowl" event with a fluency twice that of GRB (910503).

Particularly intriguing is that the highest energy particles are emitted with delays of tens of minutes. It is possible that the delayed photons observed in events like GRB 940217, where a 25~GeV photon arrived with a delay of 77~minutes, are the decay products of neutral pions photoproduced by accelerated protons interacting with the MeV photons in the GRB shock. This assumption is sufficient to compute the number of high energy neutrinos produced in the shock.

From the observed (and theoretically expected) high energy GRB gamma ray spectrum
\begin{equation}
\phi_\gamma = E_\gamma {dN_\gamma\over dE_\gamma} \sim E_\gamma^{-1} \,,
\end{equation}
we can fix the photon flux by normalizing to 1 event above 25~GeV, per EGRET area of 0.1~m$^2$ and per burst. We obtain
\begin{equation}
\phi_\gamma = {250\over E_\gamma} \rm\ GeV\ m^{-2}\ burst^{-1} \,.
\end{equation}
Pion photoproduction is dominated by the $\Delta$ resonance. Particle physics relates the neutrino and gamma ray luminosities by Eq.~(8). This is a conservative estimate because it assumes no high energy gamma ray absorption in the source. Eqs.~(8) and (19) determine the neutrino flux.

The probability to detect a neutrino can now be computed from (14) and (15). For energies below $10^3$~GeV, where both the range and cross section depend linearly on energy, $n=2$ and $A \simeq10^{-12}$. Above this energy we simply assume $n=0$, which is adequate for this calculation. The number of detected events is obtained by folding the neutrino flux with the detection probability 
\begin{equation}
N_\nu = \int_{E_{\nu\,\rm th}} {dN_\nu\over dE} P_{\nu\to\mu}(E) dE\,.
\end{equation}
Here $E_{\nu\, th}$ is the threshold of the detector. For the flux of Eq.~(19) (multiplied by 3/13) we obtain
\begin{equation}
N_\nu = {3\over 13} 250 \int_{E_\nu\, th} {dE\over E^2} P_{\nu\to\mu}(E)\,.
\end{equation}
Therefore
\begin{equation}
N_\nu = 5\times10^{-8}\rm\ m^{-2}\ burst^{-1} \simeq 10\ km^{-2}\ year^{-1}\,.
\end{equation}
For 1 burst per day one predicts more than 10 events per squared kilometer per year. These events are relatively easy to detect because they come at precise times from known directions, supplied by the gamma ray observations. One  may detect that many events from a single burst that is 10 times closer, and, possibly, a handful of events in the operating AMANDA detector with an area of $\sim 10^4 \rm\ m^2$. Notice that the muons are near TeV energy provided the detector threshold is below this value, as is the case for AMANDA with a threshold of $\sim50$~GeV.

The calculation can be done in a slightly different way, yielding the same result. We can think of the GRB phenomenon as a source of energy releasing a sizeable fraction of a solar mass every day. This corresponds to a flux of $10^{47} \sim 10^{48}$~ergs/sec. For an $E^{-2}$ spectrum Eq.~(20) can be conveniently rewritten in the form
\begin{equation}
N_\nu = 10 \left[ {\cal L}_\nu \over 10^{38}{\rm\ erg/s} \right]
\left[ 10{\rm\ kpc} \over d \right]^2 \left(\rm 10^4\, m^2\,year\right)^{-1}\,,
\end{equation}
i.e., 10 neutrinos of TeV energy are observed in an AMANDA-size detector per year for a galactic source with Eddington luminosity. For the GRB sources the flux is increased by almost ten orders of magnitude, though the distance $d \simeq 3 \times 10^3$~Mpc ($\left<z\right>\sim1$) is more than $10^5$ times larger. The increased luminosity does not quite compensate for distance to the sources and Eq.~(23) yields $10 \sim 10^2$ events per year in a kilometer squared detector. This confirms our previous estimate.

In the previous estimates we have obviously assumed that kinetic energy in the shock is efficiently converted into the acceleration of protons, an assumption intrinsic to models where GRBs are the sources of the highest energy cosmic rays. Our calculations assume that pions are the source of the observed gamma rays. While such models\cite{Pac} may be somewhat disfavored, Waxman and Bahcall\cite{bahcall} have pointed out that the production of high energy neutrinos may be a feature of more conventional models as well. In their model the gamma rays are produced by synchrotron radiation by relativistic electrons accelerated in the shock, possibly followed by inverse-Compton scattering. Provided that the efficiency with which kinetic energy is converted to accelerated protons is comparable to that for electrons, the GRBs can still generate the highest energy cosmic rays\cite{waxman}. These protons will photoproduce pions and neutrinos on the MeV gamma rays in the burst. The predicted rates are similar to the ones estimated above, but the neutrinos are some two orders of magnitude higher in energy. The predicted neutrino flux is 
\begin{eqnarray}
	dN/dE &=& A / E^2  \hskip3.5em       {\rm for}\  E > E_b\\
	      &=& A / (E_b E) \qquad    {\rm for}\   E < E_b
\end{eqnarray}
with $A = 4 \times 10^{-13}$~TeV~(cm$^2$ s sr)$^{-1}$ and $E_b \simeq 100$~TeV.

The flux is similar to the one obtained for AGN in the previous section. This is not surprising as the mechanisms are similar. It is higher than the AGN flux, but most neutrinos are produced near the maximum energy of 100~TeV, rather than 100~PeV. The event rates are in the end similar.

There is also a very different class of models where the delayed high energy gamma rays are produced in the  interactions of fireball protons with hydrogen gas clouds\cite{katz}. For the  case of photoproduction, pions are photoproduced on the $\Delta$ resonance and textbook particle physics implies that the neutrino luminosity is 3/13 of the gamma ray luminosity. For this alternative model where the gamma rays originate in nuclear interactions, the luminosities of photons and neutrinos should be roughly equal, and, if anything, the neutrino luminosity larger. The question is whether absorption effect (of gammas in the source) are small enough for these predictions to survive.
  
The association of GRBs and high energy cosmic rays has other powerful observational implications. Models predict fluxes of high energy photons and neutrinos which exceed cosmic ray proton fluxes by a couple orders of magnitude and I guess that, in the near future, air Cherenkov telescopes, air shower arrays, neutrino telescopes and, of course, the 10000~km$^2$ ``Auger" cosmic ray detector will begin to probe this intriguing association.

\newpage
\section*{Acknowledgements}

A lot of the material in this lecture is based on work with Enrique Zas. This work was supported in part by the University of Wisconsin Research Committee with funds granted by the Wisconsin Alumni Research Foundation, in part by the U.S.~Department of Energy under Grant No.~DE-FG02-95ER40896, and in part by the CICYT under contract AEN96-1773.

\end{document}